\documentclass{IEEEtran}
\usepackage{cite}
\usepackage[utf8]{inputenc}
\usepackage{amsmath}
\usepackage{amssymb}
\usepackage{graphicx}
\usepackage{url}
\newcommand{\cdummy}{\cdot}
\newcommand{\nobracket}{}

\newcommand{\tmem}[1]{{\em #1\/}}

\begin{document}

\title{Far-field compressive ultrasound beamforming}
\author{Nikunj Khetan, Jerome Mertz
\thanks{This work was partially supported by the National Institutes of Health (R21GM134216) and by the Boston University Photonics Center.}
\thanks{N. K. is with Boston University Mechanical Engineering, 110 Cummington Mall, Boston, MA 02215 (e-mail: nkhetan@bu.edu). }
\thanks{J. M. is with Boston University Biomedical Engineering, 44 Cummington Mall, Boston, MA 02215 (e-mail: jmertz@bu.edu).}}
	
\maketitle

\begin{abstract}
We present a compressive beamforming method for coherent plane-wave compounding (CPWC) ultrasound imaging based on a far-field decomposition of the received  radiofrequency (RF) data into virtual plane waves. This decomposition recasts the imaging operation entirely in the spatial frequency domain ($k$-space), allowing direct and flexible control over $k$-space sampling distributions based on the principle of coarrays. We present vernier-type sampling strategies designed to optimize the tradeoff between image contrast and resolution with minimum redundancy, including strategies that favor dense low-frequency sampling for high contrast, shifted schemes that extend the frequency support for improved resolution, and confocal or hybrid compounding schemes that approximate the spatial-frequency transfer function of conventional DAS beamforming. Our method, called KK beamforming, is validated with a calibration phantom and in-vivo human tissue data, demonstrating compression factors of an order of magnitude while maintaining image qualities comparable to conventional DAS. We further demonstrate that KK beamforming yields improvements in computational speed owing to its reduced memory footprint and more efficient cache utilization of the compressed data and associated look-up tables.
\end{abstract}            

\section{Introduction}

Ultrasound image formation is most commonly performed using delay-and-sum (DAS) beamforming, because of its simplicity and robustness. In conventional DAS imaging \cite{thomenius_evolution_1996,jensen_ultrasound_2002,perrot_so_2021} , acoustic pulses are launched into a sample as narrow beams, and resultant reflected radiofrequency (RF) signals are received across a transducer array. The receive signals are delayed and coherently summed to obtain a focus at each sample point. While DAS is favored for its computational simplicity and robustness, its performance is intrinsically tied to the density of the sampled channel data, necessitating large storage and high-bandwidth architectures for high-resolution imaging. This has motivated the development of methods to reduce data acquisition requirements without significantly compromising image quality. Examples include compressive beamforming approaches in the temporal and frequency domains that exploit structure in the received signals, such as finite rate of innovation \cite{wagner_compressed_2012}. DAS signals in these approaches are represented in the temporal Fourier domain and reconstructed from a subset of Fourier coefficients  \cite{chernyakova_fourier-domain_2014}. However, temporal frequency compression alone does not exploit redundancy across the array, leaving the spatial dimension largely uncompressed.

Spatial compression methods directly target redundancy across array elements. Examples of such methods have a long history generally revolving around the concept of coarrays \cite{hoctor_unifying_1990}. In transmit/receive geometries with $N$ transmit elements and $M$ receive elements, in principle $N \times M$ degrees of freedom are available to extract information from the sample, leading to the concept of an effective aperture \cite{lockwood_optimizing_1996}. For a linear array these degrees of freedom can be distributed to minimize redundancy \cite{moffet_minimum-redundancy_1968, lockwood_optimizing_1996, mitra_general_2010}, thus optimizing information content and enabling acceptable imaging even with relatively few actuator elements. Further reduction in the number of physical actuator elements can be obtained with compressive sensing (CS) strategies, for example leveraging assumptions on sample sparsity and/or signal randomness. CS strategies have been implemented in combination with MIMO radar \cite{rossi_spatial_2014}, wave-atom signal representation  \cite{liebgott_pre-beamformed_2013}, minimum variance beamforming  \cite{paridar_plane_2023}, to name a few. 

To achieve ultrafast imaging, DAS has been extended by the method of coherent plane-wave compounding (CPWC) wherein multiple plane waves are transmitted at different angles and their echoes are coherently summed \cite{montaldo_coherent_2009}. CPWC dramatically increases frame rates compared to conventional focused transmit schemes while recovering image quality through angular compounding, making it particularly relevant for ultrafast imaging applications such as cardiac \cite{liu_high_2022}, vascular \cite{martinoli_power_1998,van_der_ven_high-frame-rate_2017}, or functional ultrasound \cite{tang_functional_2020,liu_time-lagged_2024}. However, this increase in speed comes at the cost of larger data volumes: each plane-wave angle generates full channel datasets across the aperture. As a result, CPWC amplifies the memory, transfer, and processing demands inherent in DAS, further motivating the development of compressive beamforming methods. Several of the aforementioned approaches developed for conventional DAS have also been applied to CPWC, such as sub-Nyquist sampling in the temporal frequency domain \cite{cohen_fourier_2015} and the implementation of CS strategies \cite{liu_compressed_2017,chernyakova_fourier-domain_2018}.

In our present work, we revisit some of the spatial compression approaches described above and apply these to CPWC. An advantage of CPWC is that the transmission of plane waves readily lends itself to a spatial frequency, or $k$-space, interpretation \cite{lambert_ultrasound_2022}. The same cannot be said of the receive component of CPWC, which involves a summation over elements in the array space. Because the transducer array is in the near field, its $k$-space interpretation is spatially variant and not as straightforward. Fourier beamforming algorithms have been developed (e.g. \cite{jian-yu_lu_2d_1997}) involving 2D FFTs of the RF receive data followed by interpolation. A simpler approach, however, involves the decomposition of the receive data directly into plane waves analogous to the transmit plane waves. This decomposition can be performed by a simple temporal shearing operation applied to the RF receive data, enabling the receive data to also be directly interpreted in $k$-space \cite{kruizinga_plane-wave_2012}, thus transposing the aperture fully into the far field for both transmission and reception. We can then bring to bear the machinery of coarrays in $k$-space rather than aperture space. In particular, we describe here different vernier-type \cite{lockwood_optimizing_1996} plane-wave decomposition algorithms applied to the RF data leading to dense spatial-frequency supports that can be tuned to favor image contrast or resolution, or a hybrid coherent/incoherent combination of both. Distributed vernier-type approaches are also implemented to mimic the frequency support of conventional CPWC beamforming. We refer to these algorithms as KK beamforming, and demonstrate their capacity to compress RF data by an order of magnitude while preserving image quality comparable to CPWC DAS. Demonstrations are performed with both calibrated phantom and human tissue imaging. 

\section{METHODS}

\subsection{Frequency support}

Throughout this work, we consider beamforming with a conventional linear transducer array. The resulting 2D amplitude image is given by \cite{lambert_ultrasound_2022}
\begin{equation}
  \label{rspace} B (\mathbf{r}) = \int H_{out} (\mathbf{r}-\mathbf{r}')
  O (\mathbf{r}') H_{in} (\mathbf{r}' -\mathbf{r}) d^2 \mathbf{r}'
\end{equation}
where $\mathbf{r}= \{ x, z \}$ is the 2D image coordinate, $O$ is the sample reflectivity (object function), and $H_{in}$ and $H_{out}$ are respectively the input (transmit) and output (receive) amplitude point spread functions, constructed computationally. Equation \ref{rspace} can be recast in Fourier space as
\begin{equation}
  \label{kspace} \hat{B} (\mathbf{k}) = \iint \hat{H}_{out}
  (\mathbf{k}_o)  \hat{H}_{in} (\mathbf{k}_i)  \hat{O} (\mathbf{k})
  \delta^2 (\mathbf{k}+\mathbf{k}_i -\mathbf{k}_o) d^2 \mathbf{k}_i d^2
  \mathbf{k}_o
\end{equation}
where $\mathbf{k}= \{ k_x, k_z \}$ is the 2D spatial frequency coordinate ($\hat{B} (\mathbf{k}) = FT [B (\mathbf{r})]$, etc.), and $\delta^2$ is the 2D delta function. The interpretation of Eq. \ref{kspace} is clear. The spatial frequencies recovered from the object by beamforming are given by $\mathbf{k}=\mathbf{k}_o -\mathbf{k}_i$, corresponding to the frequency support associated with the beamforming operation. To avoid aliasing, this frequency support should ideally be continuous, as suggested by the integrals in Eq. \ref{kspace}. In practice, however, the frequency support is discrete and care must be taken in how the spatial frequencies $\mathbf{k}$ are sampled. For example, to avoid aliasing in the transverse direction, $k_x$ should be sampled with a step size $\delta k_x$ no larger than $S^{- 1}$, where $S$ is the transverse spatial extent of the recovered object. This is essentially the Nyquist criterion in the spatial frequency domain where we require a sampling rate of $2/S$ to avoid aliasing. In the case considered here of imaging with a linear transducer array, $S$ is generally defined as the full aperture size of the array.

Our goal is to formulate an approach to compress the RF data required for beamforming while at the same time sampling $\mathbf{k}$ as densely and non-redundantly as possible. For this, we draw guidance from the principle of CPWC, which involves the insonification of the sample with a sequence of $N$ plane waves of off-axis tilt angles $\theta_i$, where $i = 1, \ldots, N$. A key advantage of CPWC is that it provides direct control over $\mathbf{k}_i$. Indeed, a spatial frequency can be directly associated with each transmit plane wave, given by $\mathbf{k}_i = \nu \mathbf{s}_i / c$, where $\nu$ is the plane-wave temporal frequency (approximated as the transducer center frequency), $c$ is the speed of sound in the object medium (assumed constant), and $\mathbf{s}_i
= \{ s_{i x}, s_{i z} \} = \{ \sin \theta_i, \cos \theta_i \}$ are unit directional vectors (note that even though $\mathbf{s}_i$ are two-element vectors, they are parametrized by a single-element scalars $\theta_i$). We thus have $\hat{H}_{in} (\mathbf{k}) = \sum_i
\delta^2 (\mathbf{k}-\mathbf{k}_i)$, where we neglect apodization factors in the interest of simplicity, though these could be included in a straightforward manner. 

We emphasize here that $\hat{H}_{in}
(\mathbf{k})$ is independent of the spatial beamforming coordinate $\mathbf{r}$. That is, $\hat{H}_{in} (\mathbf{k})$ is spatially invariant. In conventional CPWC, the same simplicity does not apply to $\hat{H}_{out}$, since the latter cannot be decomposed into spatial frequencies in such a straightforward manner. In particular, the spatial-frequency decomposition of $\hat{H}_{out}$ depends on the beamforming coordinate $\mathbf{r}$, meaning it is not spatially invariant. As a result, conventional CPWC DAS beamforming is generally performed in the spatial domain rather than the Fourier domain, by performing summations over spatiotemporal hyperbolas in the RF data with curvatures that depend on $\mathbf{r}$ \cite{perrot_so_2021}.

\subsection{KK beamforming}

\begin{figure*}
\centering
\includegraphics[width= \textwidth]{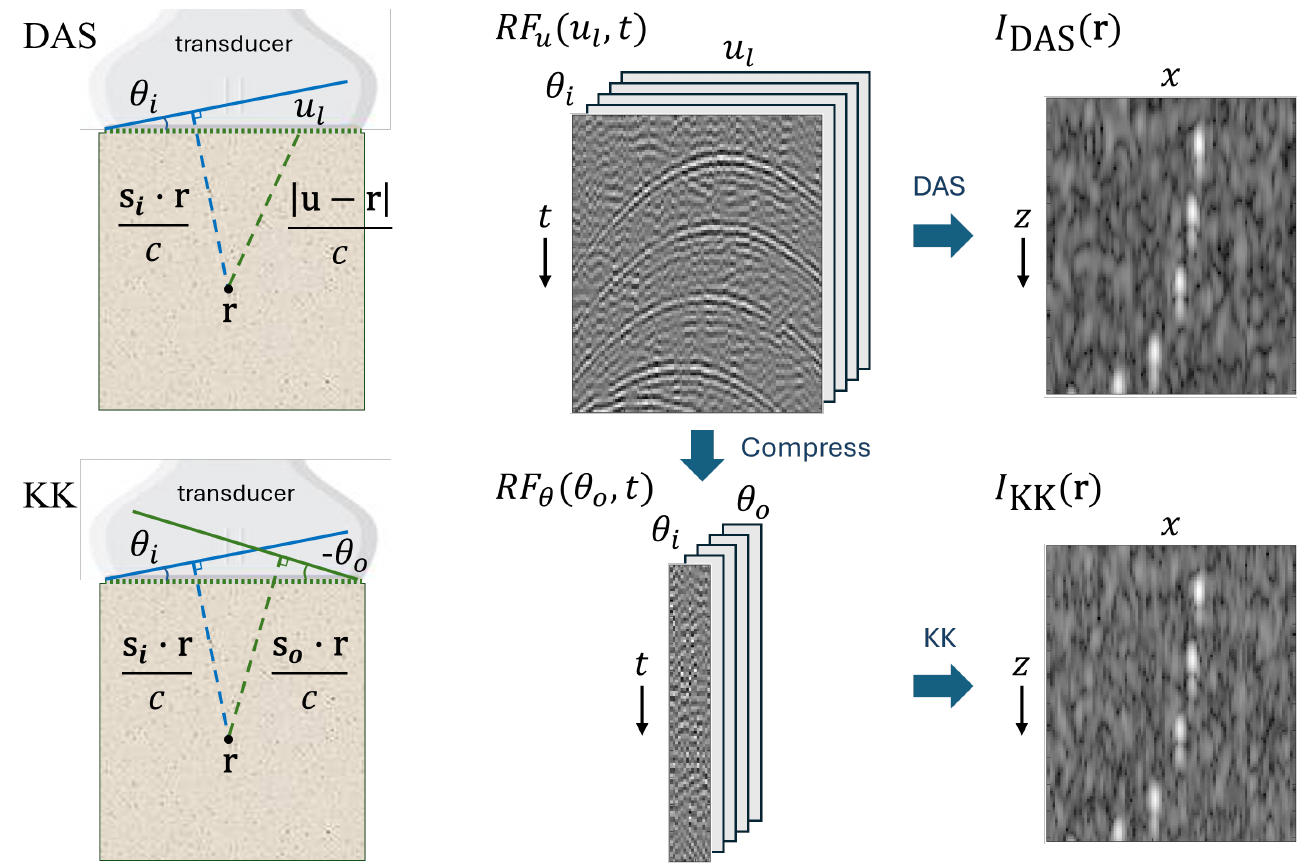}
\caption{Top row: Conventional CPWC DAS beamforming. For every transmit angle $\theta_i$, the received RF data is a function of $u_l$ and $t$. Time delays to sample coordinate $\mathbf{r}$ are shown in left panel. Bottom row. KK beamforming. RF data is first compressed (Eq. \ref{compression}) and then beamformed (Eq. \ref{KK}) using time delays to sample coordinate $\mathbf{r}$ shown in left panel.}
\label{fig1}
\end{figure*}

We adopt here an alternative beamforming approach that is fully spatially invariant. In effect, this approach is a symmetrized version of CPWC where both transmit and receive signals are plane waves (see Fig. 1). Our approach involves two steps. The first is a compressive pre-processing step where the raw RF data is recast in the spatial frequency domain. The raw RF data is given by $RF_u(u_l,t)$, where $u_l$ is the transducer coordinate and $t$ is time delay. This is transformed by the operation
\begin{equation}
  \label{compression} RF _{\theta}(\theta_o, t) \leftarrow \sum_{l =
  1}^L RF_u (u_l, t + d_l \sin \theta_o / c)
\end{equation}
where $d_l = u_l$ for positive $\theta_o$ and $d_l = u_l - u_L$ for negative $\theta_o$. This operation corresponds to first shearing $RF_u
(u_l, t)$ in time and then summing along the horizontal spatial direction, both of which are computationally inexpensive. Note that prior to this operation, the raw RF data is a matrix of size $L \times T$, where $T$ is the total acquisition time (both $L$ and $T$ are normalized here to the transducer spatiotemporal sampling resolution, meaning both are integers). After the operation defined by Eq. \ref{compression}, the RF matrix is of size $M
\times T$, where $M$ is the total number of synthesized receive angles, defined by the user. The compression factor is thus given by $L / M$, which, as shown below, can be chosen to reduce the data size typically by an order of magnitude.

Our second step involves a modified version of DAS beamforming. Conventionally, CPWC DAS beamforming is defined by
\begin{equation}  
  \label{DAS} B_{DAS} (\mathbf{r}) = \sum_{\theta_i, u_l} RF_u
  (u_l, \tau_{in} (\mathbf{r}, \theta_i) + \tau_{out}
  (\mathbf{r}, u_l))
\end{equation}
where $\tau_{in} (\mathbf{r}, \theta_i) =\mathbf{s}_i \cdummy
\mathbf{r}/ c$ is the input (transmit) delay, and $\tau_{out}
(\mathbf{r}, u_l) = | \mathbf{r}- \{ u_l, 0 \} | / c$ is the output (receive) delay, with the transducer assumed to be located at $z = 0$. We adopt instead the modified beamforming approach called KK beamforming, defined by
\begin{equation}
  \label{KK} B_{KK} (\mathbf{r}) = \sum_{\theta_i, \theta_o} RF_{\theta}
  (\theta_o, \tau_{in} (\mathbf{r}, \theta_i) + \tau_{out}
  (\mathbf{r}, \theta_o))
\end{equation}
where $\tau_{in} (\mathbf{r}, \theta_i) =\mathbf{s}_i \cdummy
\mathbf{r}/ c$ is the input (transmit) delay, as before, and $\tau_{out} (\mathbf{r}, \theta_o) =\mathbf{s}_o \cdummy \mathbf{r}/ c$ is the output (receive) delay. Again, for simplicity we have neglected apodization factors in both Eqs. \ref{DAS} and \ref{KK}, though these could be included in a straightforward manner. We emphasize that the summation in Eq. \ref{KK} is entirely independent of $u_l$. $RF_{\theta}(\theta_o, t)$ may thus be considered fully in the far field, leading to $\hat{H}_{out}
(\mathbf{k}) = \sum_o \delta^2 (\mathbf{k}-\mathbf{k}_o)$. We note also that in the ideal case of fine sampling with a wide bandwidth transducer, Eqs. \ref{DAS} and \ref{KK} are equivalent to Eq. \ref{rspace} \cite{lambert_ultrasound_2022}. In the more realistic case where the sampling of $\mathbf{k}_i$ and $\mathbf{k}_o$ is discretized we must take care to avoid aliasing, particularly when the RF data is highly compressed. We turn now to simple approaches to mitigate such aliasing based on the concept of coarrays.

As noted above, the frequency support associated with beamforming is defined by $\mathbf{k}=\mathbf{k}_o -\mathbf{k}_i \simeq \nu (\mathbf{s}_o-\mathbf{s}_i) / c$. When performing CPWC with KK beamforming, $\mathbf{k}_i$ and $\mathbf{k}_o$ are thus directly controllable by the choice of input and output plane waves. The former are physical plane waves used to insonify the sample; the latter are virtual plane waves synthesized from the receive RF data. Conventionally, the tilt angles associated with the transmit plane waves are uniformly distributed between $-
\theta_{\max}$ and $\theta_{\max}$, where $\theta_{\max}$ is the maximum tilt angle supported by the transducer $f$-number. While the assumption of uniformly distributed transmit angles is not a requirement here, we adopt it so that our approach can be applied to standard CPWC data, without any modifications imposed on conventional transmit protocols. The angular spacing between transmit angles is thus $\delta \theta_i = 2 \theta_{\max} / (N - 1)$, leading to a spacing in the transverse component of $k_i$ given by $\delta
k_{i x} = \nu \delta s_{i x} / c$  (for paraxial angles we have $\delta s_{i x}
\simeq \delta \theta_i$). Our problem therefore reduces to determining a set of receive angles $\theta_o$, under the constraints of a prescribed $\theta_{\max}$ and desired $M$ (i.e. a prescribed maximum angular range and desired compression factor), such that $k_x$ remains as densely sampled as possible. Several approaches can be adopted for this, of which we detail only a few.

\subsection{Contrast versus resolution}

\begin{figure*}[!]
\centering
\includegraphics[width= \textwidth]{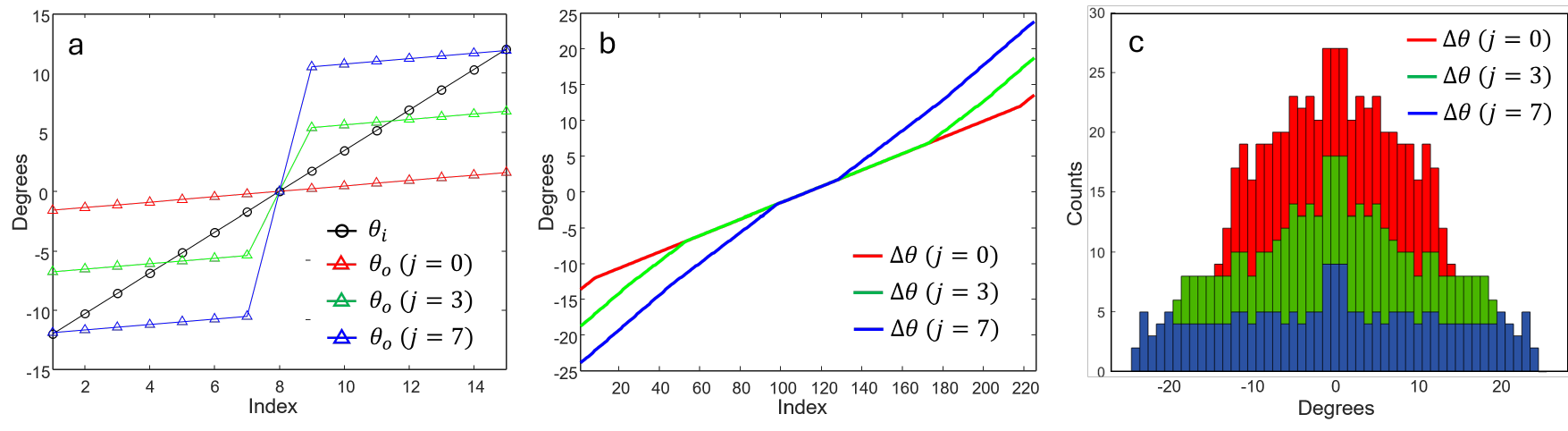}
\caption{Examples of KK frequency support. a) Transmit angles $\theta_i$ and (sorted) receive angles $\theta_o$ for different shift parameters $j$, where angular range is $\pm 12^{\circ}$ for both and $N=M=15$. b) Resultant distribution of difference angles $\Delta \theta = \theta_o - \theta_i$, (proportional to $\Delta k_x$ in the paraxial limit). c) Stacked histograms of $\Delta \theta$.    }
\label{fig2}
\end{figure*}

To begin, it is instructive to consider an approach that is {\tmem{not}} optimal. At first glance, one might consider beamforming with the same set of tilt angles both in the transmit and receive directions. This is a poor strategy because, although $k_x$ spans twice the range of $k_{i x}$ and $k_{o x}$ individually (leading to a doubling of frequency support), the sampling of $k_x$ becomes highly redundant with large steps of size $\delta k_x = \delta k_{i x} =\delta k_{o x}$, leading to severe aliasing unless $N = M$ is large (approaching $L$). In other words, this strategy is not amenable to data compression.

A better strategy involves selecting a distribution of $\theta_o$'s such that they uniformly ``fill the gap'' between $\theta_i$'s, in analogy with vernier interpolation \cite{lockwood_optimizing_1996}. That is, we consider the sequence $\theta_o = 2 o \delta \theta_i / M$ with $o = - \lfloor M / 2
\rfloor, \ldots, \lfloor M / 2 \rfloor$, where $\lfloor \ldots \rfloor$ indicates a truncation towards zero. The sampling step size then becomes $\delta k_x \simeq 2 \delta k_{i x} / M$, corresponding to a reduction in step size by a factor of $M/2$ compared to the example above. Thus, even in cases where $M$ is much smaller than $L$, it can be chosen large enough to achieve the condition $\delta k_x \ll
\delta k_{i x}$, thus achieving high sampling density. This simple strategy reduces the possibility of aliasing and thus provides high contrast (see Results). On the other hand, it fails to provide high resolution since the transverse frequency support $\Delta k_x$, defined as the full range of $k_x$ sampling, is now given by $\Delta k_x = \Delta k_{i x} (1 + 2 / N)$. That is, rather than being doubled as in the example above, the frequency support becomes only marginally increased from that provided by the transmit angles alone.

To extend the range of $\Delta k_x$ we can continue to ``fill the gap'' between $\theta_i$'s by using the same sequence of $\theta_o$'s as defined above, but this time applying a shift to this sequence such that it is no longer centered about $\theta_o = 0$ (see Fig. \ref{fig2}). To maintain symmetry in the distribution of $\theta_o$'s, this involves applying opposing shifts for positive and negative $\theta_o$'s. That is, we perform KK beamforming using the sequence of $\theta_o$'s defined by
\begin{equation} \label{uniform}
\begin{split}
  \theta_o &= \text{sgn}(o) \delta \theta_i  (2 | o | / M + j) \\
  &o = - \lfloor M / 2 \rfloor, \ldots, \lfloor M / 2 \rfloor
\end{split}
\end{equation}
where $j$ is an integer between 0 and $\lfloor N / 2 \rfloor$. In the case $j
= 0$, we retrieve the same results as above. In the case $j = \lfloor N / 2
\rfloor$, we obtain a sampling range $\Delta k_x$ that is roughly double that of $\Delta k_{i x}$, leading to roughly a doubling of the frequency support. On the other hand, the sampling density becomes commensurately reduced, with a sampling step size now given by $\delta k_x \simeq 4 \delta k_{i x} / M$, leading to increased susceptibility to aliasing and production of sidelobes. Of course, intermediate values of $j$ may be chosen. These lead to a sampling of $k_x$ that is no longer uniformly distributed but rather bimodal, with denser sampling when $|k_x| \lesssim 
 (N/2-j)k_{ix}$ and sparser sampling beyond this range. Such intermediate values can provide a satisfactory compromise between contrast and resolution. In addition, beamformed images $B^{(j)}_{KK} (\mathbf{r})$ obtained from different values of $j$ can be compounded to improve overall image quality. This compounding can be coherent or incoherent, yielding final image intensities given by
\begin{equation} \label{coherent}
  I^{(coh)}_{KK} (\mathbf{r}) = \left| \sum_j
  B^{(j)}_{KK} (\mathbf{r})  \right|^2
\end{equation}
\begin{equation} \label{incoherent}
  I^{(inc)}_{KK} (\mathbf{r}) = \sum_j | B^{(j)}_{KK}
  (\mathbf{r}) |^2
\end{equation}
We note that the values of $j$ in these summations are drawn from the interval 0 and $\lfloor N / 2 \rfloor$, and need not be sequential. Such a hybrid coherent/incoherent approach requires $J$ beamforming operations. The resultant effective data compression factor thus becomes $L / (J \times M)$, which, nevertheless, can remain significantly larger than 1 (see Results). Examples of hybrid coherent/incoherent imaging have been performed before (e.g. \cite{lu_improving_2018}), though to our knowledge not in the context of far-field KK beamforming.

\subsection{Confocal KK beamforming}

\begin{figure*}[!]
\centering
\includegraphics[width= \textwidth]{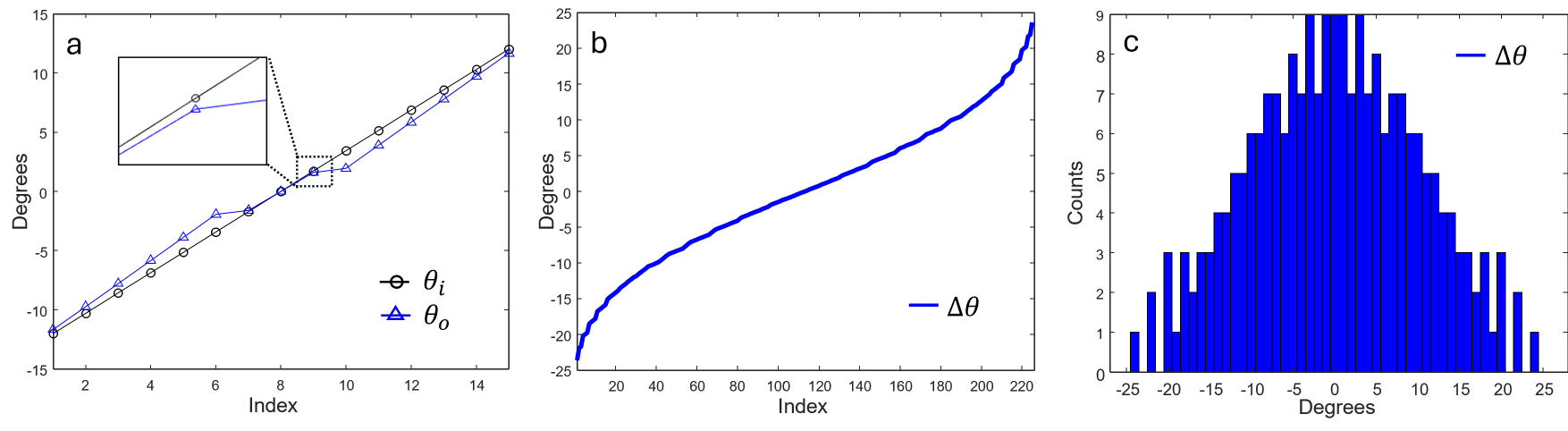}
\caption{Example of KK frequency support. a) Transmit angles $\theta_i$ and (sorted) receive angles $\theta_o$, where angular range is $\pm 12^{\circ}$ for both and $N=M=15$. b) Resultant distribution of difference angles $\Delta \theta = \theta_o - \theta_i$, (proportional to $\Delta k_x$ in the paraxial limit). c) Histogram of $\Delta \theta$ closely resembles confocal-like frequency transfer of conventional CPWC DAS.    }
\label{fig3}
\end{figure*}

From Eqs. \ref{rspace} and \ref{kspace}, we observe that ultrasound beamforming corresponds to a confocal imaging operation, with an overall amplitude point spread function given by $H_{conf} (\mathbf{r}) =
H_{out} (-\mathbf{r}) H_{in} (\mathbf{r})$ and frequency transfer function given by $\hat{H}_{conf} (\mathbf{k}) =
\hat{H}_{out} (\mathbf{k}) \ast \hat{H}_{in} (\mathbf{k})$, where $\ast$ indicates a convolution. When performing conventional DAS beamforming in the ideal case where the transducer elements are small and $H_{out}$ and $H_{in}$ share the same aperture, then $|
\hat{H}_{out} (k_x) | = | \hat{H}_{in} (k_x) |$ become simple rectangle functions (neglecting apodization and obliquity factors), and $\hat{H}_{conf} (k_x)$ becomes a simple triangle function. As shown in Fig. \ref{fig2}c, such transfer function behavior can be readily mimicked with KK beamforming by performing compounding of multiple beamformed images obtained with different shift values $j$. Alternatively, we can adopt a generalization of the multiple-beamforming method such that the full range $j$ values becomes distributed within a single beamforming operation. That is, we perform KK beamforming using the sequence of $\theta_o$'s defined by 
\begin{equation}
\begin{split}
  \theta_o &= \text{sgn} (o) \delta \theta_i  (2 | o | / M + \text{mod}
  (\nobracket | o |, \lfloor N / 2 \rfloor \nobracket))  \\
&o = - \lfloor M
  / 2 \rfloor, \ldots, \lfloor M / 2 \rfloor
\end{split}
\label{confocal}
\end{equation}

As observed in Fig. \ref{fig3}b, the resulting distribution of $k_x$ sampling is no longer bimodal. It remains denser near the origin, as before, but now decreases in density in a continuous manner away from the origin. A histogram of the $k_x$ sampling values reveals that the resulting sampling density closely resembles a triangle (Fig. \ref{fig3}c). As such, ${I
_{KK}}^{(conf)} (\mathbf{r}) = | B _{KK} (\mathbf{r})
|^2$ obtained from such beamforming is expected to closely resemble an image obtained from conventional DAS beamforming, despite being constructed from highly compressed RF data (with data compression factor $ L/M$). We show below that very acceptable imaging results can be obtained with data compression factors as large as an order of magnitude.

\subsection{Experimental Methods}

Results presented here were all acquired using a Verasonics Vantage 256 system (Kirkland, WA, USA) connected to a GE9LD linear array probe. The phantom experiments were conducted on a CIRS 040 GSE phantom. The human imaging data was collected from a healthy volunteer according to IRB protocol \#5914E, approved by Boston University. Key parameters for each experiment are described in Table \ref{tableMethods}. Data processing was performed in Matlab (The MathWorks Inc., Natick, MA, USA) on a Dell Precision Tower 5810 with an Intel Xeon E5-2680 v3 processor. The host controller (desktop) that performed the processing had 128 GB of DDR4-2666MHz RAM. For balanced and fair comparison of our beamforming results, all images shown throughout this work were gamma compressed according to the contrast matching procedure described in Ref. \cite{khetan_plane_2025}. We also evaluated image quality using the generalized contrast-to-noise ratio (gCNR) metric described in Refs. \cite{rodriguez-molares_generalized_2020,schlunk_methods_2023}.

\begin{table*}
    \caption{List of datasets and key acquisition parameters}
    \label{tableMethods}
    \setlength{\tabcolsep}{4pt}
    \hspace*{-20pt}
    \begin{tabular}{p{100pt}p{100pt}p{100pt}p{100pt}p{100pt}p{100pt}}
        
        Properties & Fig. 4 & Fig. 5 & Fig. 6 & Fig. 7 \\
        \hline\hline
        Probe & GE9LD & GE9LD & GE9LD & GE9LD  \\
        Element pitch (mm) & 0.230 & 0.230 & 0.230 & 0.230  \\
        Number of elements & 192 & 192 & 192 & 192  \\
        Transmit freq. (MHz) & 5.2 & 5.2 & 5.2 & 5.2  \\
        Sampling freq. (MHz) & 20.83 & 20.83 & 20.83 & 20.83  \\
        Num of Samples per Acq. & 2944 & 2944 & 2048 & 2944 \\
        Sound speed (m/s) & 1540 & 1540 & 1540 & 1540  \\
        F-number & 0.9 & 2.25 & 2.25 & 0.9  \\
        Angular range (degrees) & 48 & 48 & 48 & 48  \\
        Wavelength (mm) & 0.2957 & 0.2957 & 0.2957 & 0.2957  \\
        Lateral range (mm) & 20.5 & 14.7 & 10.1 & 43.9  \\
        Lateral pixel count & 360 & 260 & 180 & 384  \\
        Lateral pixel size (mm) & 0.057 & 0.057 & 0.057 & 0.114  \\
        Axial range (mm) & 10.2 & 9.46 & 6.51 & 32.4  \\
        Axial pixel count & 280 & 260 & 180 & 440  \\
        Axial pixel size (mm) & 0.037 & 0.037 & 0.036 & 0.074  \\
        \hline\hline
        \multicolumn{5}{p{530pt}}{The GE9LD probe was obtained from GE HealthCare (Chicago, IL, USA)}
    \end{tabular}
\end{table*}

\subsection{Computational Implementation}

To assess the performance of KK beamforming compared to conventional CPWC DAS beamforming, we implemented both in C++ using the Eigen linear algebra library \cite{eigenweb}. Most of our code was wrapped in MEX functions (MATLAB’s mechanism for calling compiled C/C++ code). We refer to these combined functions as EigenMex functions, which we used to maintain compatibility with the Matlab datasets produced by our Verasonics hardware.

In fast DAS implementations, time delays are generally pre-computed and stored as Look Up Tables (LUTs) \cite{perrot_so_2021}. We adopted the same approach here. However, in the case of CPWC DAS, a computation of time delays for every image pixel ${x,z}$ would result in LUTs of size $X \times Z \times L \times N$, which would easily overflow the CPU cache of standard computers and impose slow access times to computer memory. We thus decomposed our LUT into three separate LUTs (LUT$_{1,2,3}$) of more manageable size, associated with time delays $
x \sin\theta_i/ c$,  $z \cos\theta_i/ c$ and $\sqrt{(x - u_l)^2 + z^2}/ c $, respectively. That is, LUT$_1$ and LUT$_2$, when added, store  $\tau_{in} (\mathbf{r}, \theta_i)$ whereas LUT$_3$ stores $\tau_{out}(\mathbf{r}, u_l)$ (see Eq. \ref{DAS}), requiring a total cache size of roughly $(X+Z)\times N + (X+L) \times Z$. 

KK beamforming significantly alleviates these memory challenges. In particular, we can define four LUTs, where LUT$_{1,2}$ are the same as above, and LUT$_{3,4}$ are the receive versions of these LUTs, that is, with $\theta_i$ replaced by $\theta_o$. The resulting total required cache memory is thus only $(X+Z) \times (N+M)$, which is reduced in size not only from the replacement of $L$ with $M$ but also from the more memory efficient linear decomposition of both transmit and receive delay times. 

Finally, we note that our RF data is systematically converted to complex floating-point precision by way of a Hilbert transform, performed by implementing a temporal Fourier transform of the RF data and setting negative frequencies to zero (as is usual in most beamforming implementations). The temporal shearing required for the compression step of our KK beamforming algorithm (Eq. \ref{compression}) can thus be conveniently performed directly in the frequency domain using the Fourier shift theorem, avoiding issues related to interpolation and memory reorganization that would have resulted had the temporal shearing been performed in the time domain.   

The overall processing speeds of KK, DAS and a commercially optimized version of DAS (Verasonics, Kirkland WA) are summarized in Table \ref{tableTimes}.

\section{RESULTS } 

\begin{figure*}[!]
\centering
\includegraphics[width= \textwidth]{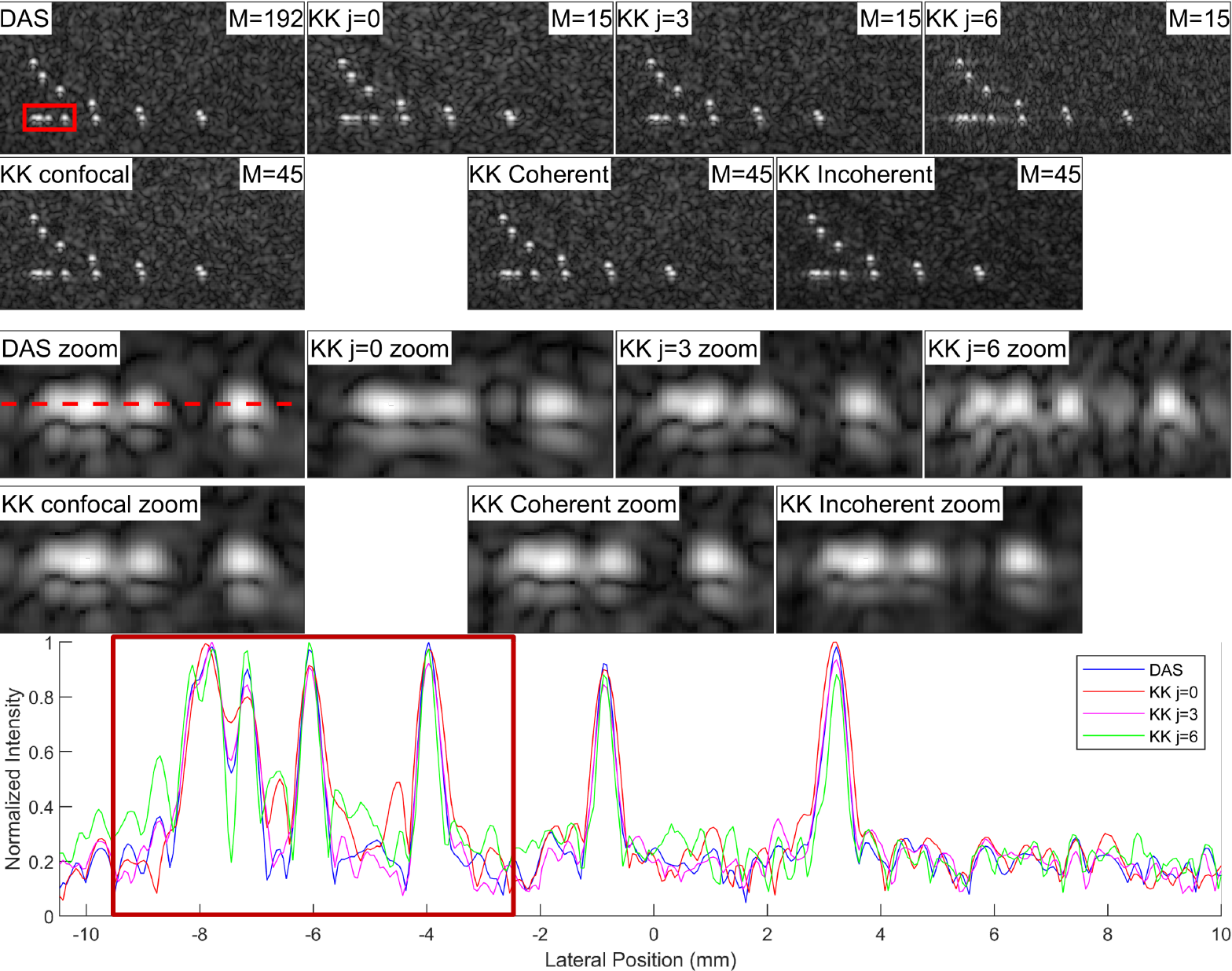}
\caption{Resolution point-target images. Top row: full field-of-view (FOV) images from DAS and KK beamforming with various receive sampling schemes. $M$ indicates number of array elements in the case of DAS or the number of synthesized receive angles in the case of KK beamforming. FOV sizes are indicated in Table \ref{tableMethods}. The red box indicates the zoomed region shown in middle rows. Middle rows: zoomed views of the closest-spaced wire targets. Bottom panel: normalized intensity profiles along red dashed line, comparing DAS with KK $j=0$, 3, 6. The $j=6$ scheme achieves improved separability of the leftmost 0.25~mm-spaced targets owing to its broader spatial frequency support, at the cost of increased susceptibility to aliasing.}
\label{figRes}
\end{figure*}

We evaluate the performance of KK beamforming using datasets obtained from resolution point targets, contrast phantoms, and human tissue. 

Results obtained from resolution point targets are shown in Fig. \ref{figRes}. We observe that, despite having significantly compressed our receive data compared to conventional CPWC DAS, KK beamforming with the various $k_o$-sampling schemes described above largely preserves the resolution and target separability achieved with DAS, though at the cost of some increased aliasing and grating lobe clutter (as becomes apparent when zooming into the region highlighted in red -- middle panels in Fig. \ref{figRes}). We note, however, that the resolution appears to actually \emph{increase} when implementing the sampling scheme from Eq. \eqref{uniform} with $j=6$. In this case, the overall frequency supports provided by DAS and KK beamforming are similarly broad, however, as illustrated in Fig. \ref{fig2}c (blue histogram), KK beamforming provides a much more uniform distribution of sampled frequencies with roughly equal weighting across all spatial frequencies. The resolution gain provided by this approach is similar to that provided by a technique of pixel reassignment\cite{sommer_pixel-reassignment_2021,sommer_k-space_2023}, though here much simpler to achieve computationally. On the other hand, $k$-space sampling with $j=6$ is sparser than that for smaller $j$'s, thus leading to more aliasing/grating lobe clutter and background noise.

\begin{figure*}[!]
\centering
\includegraphics[width= \textwidth]{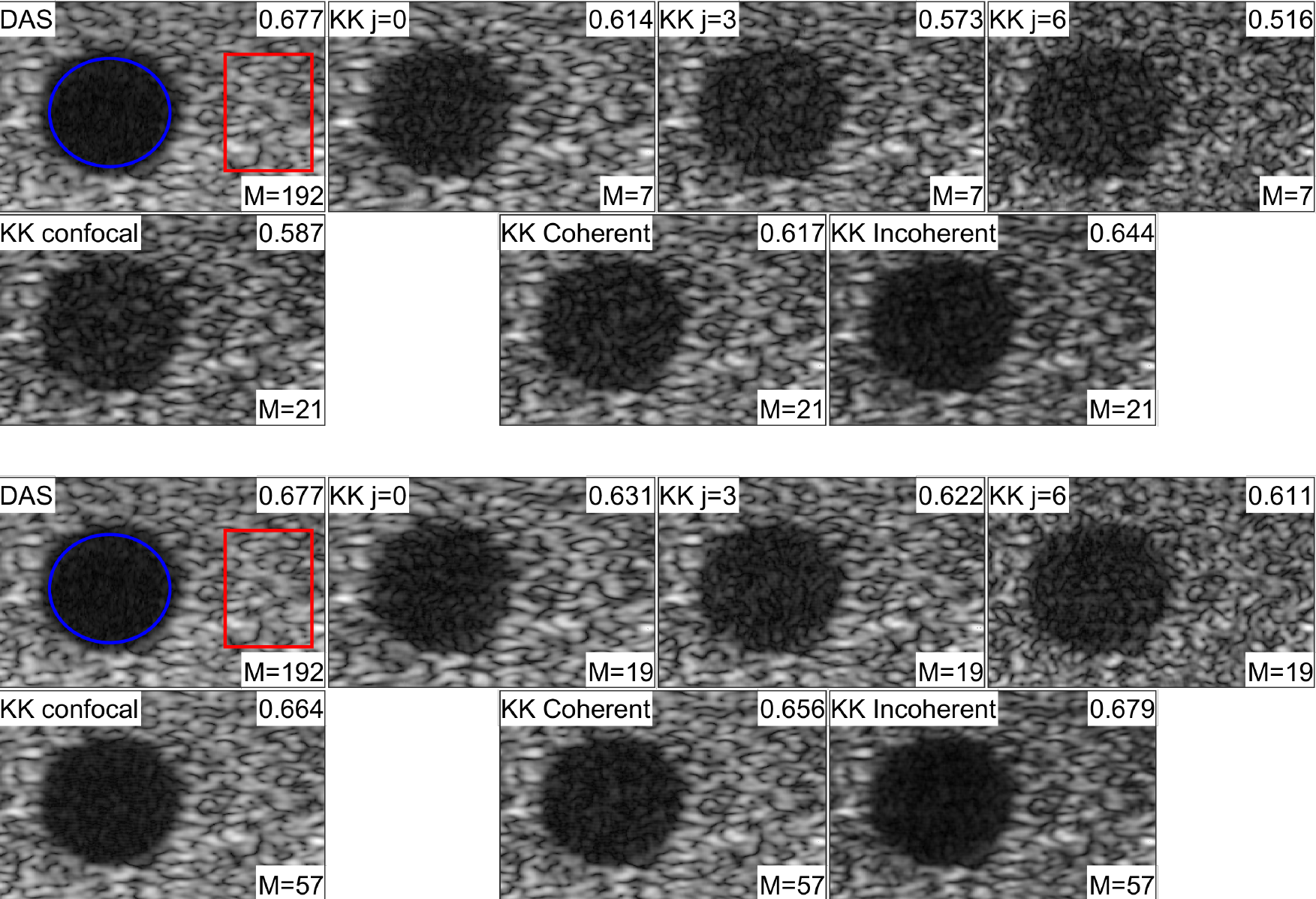}
\caption{Anechoic inclusion with associated gCNR values (inclusion defined by blue circle; background defined by red rectangle). All panels obtained with $N=15$ transmit angles; $M$ indicates the number of receive elements or angles.
Row 1 shows conventional DAS image ($M=192$) and KK images obtained with $27\times$ compressed RF data ($M=7$) using Eq. \ref{uniform}. Row 2 shows KK images obtained from confocal beamforming (Eq. \ref{confocal} -- left panel) and from compounding $j=0,3,6$ panels directly above, either coherently (Eq. \ref{coherent}) or incoherently (Eq. \ref{incoherent}). For all panels in Row 2,  $M=21$ overall . Rows 3 and 4 are the same as Rows 1 and 2, except with threefold more receive angles. The resulting gCNR values indicated in top/right of each panel are comparable to DAS for a compression factor of $3.4 \times$ ($M=57$), though somewhat deteriorate with increasing compression (decreasing $M$).}
\label{figContrast}
\end{figure*}

We can quantify the impact of KK beamforming on clutter and noise. In Fig. \ref{figContrast}  we show results from various $k_o$-sampling schemes with different total numbers $M$ of synthesized receive angles. Here, we evaluate the gCNR associated with an anechoic inclusion (delimited by blue circle). The first and third rows respectively illustrate images obtained with the $k_o$-sampling approach described by Eq. \ref{uniform} for different values of $j$. The compression factors here are significant ($27 \times$ and $10 \times$ respectively). Even with these high compression factors, the inclusion remains readily apparent. As above, we observe that larger values of $j$, while leading to broader frequency support and increased resolution, also commensurately lead to sparser spatial-frequency sampling, resulting in increased aliasing/clutter (as evidenced by a reduced gCNR). 

Rows two and four in Fig. \ref{figContrast} illustrate results obtained from still different $k_o$-sampling approaches, though with less severe compression factors ($9.1 \times$ and $3.4 \times$ respectively). The panels labeled `KK confocal' are obtained from the $k_o$-sampling approach described by Eq. \ref{confocal}, whereas the panels labeled `KK coherent' and `KK incoherent' are synthesized from the three panels directly above, using coherent (Eq. \ref{coherent}) or incoherent (Eq. \ref{incoherent}) compounding respectively. Interestingly, the hybrid approach of utilizing sub-images obtained with different $j$ values and compounding these \emph{incoherently} seems to yield the highest contrast of all. For example, the gCNR obtained from this hybrid approach actually slightly outperforms that obtained from conventional DAS even with a compression factor of $3.4 \times$ (row 4, rightmost panel).    

\begin{figure*}[!]
\centering
\includegraphics[width= \textwidth]{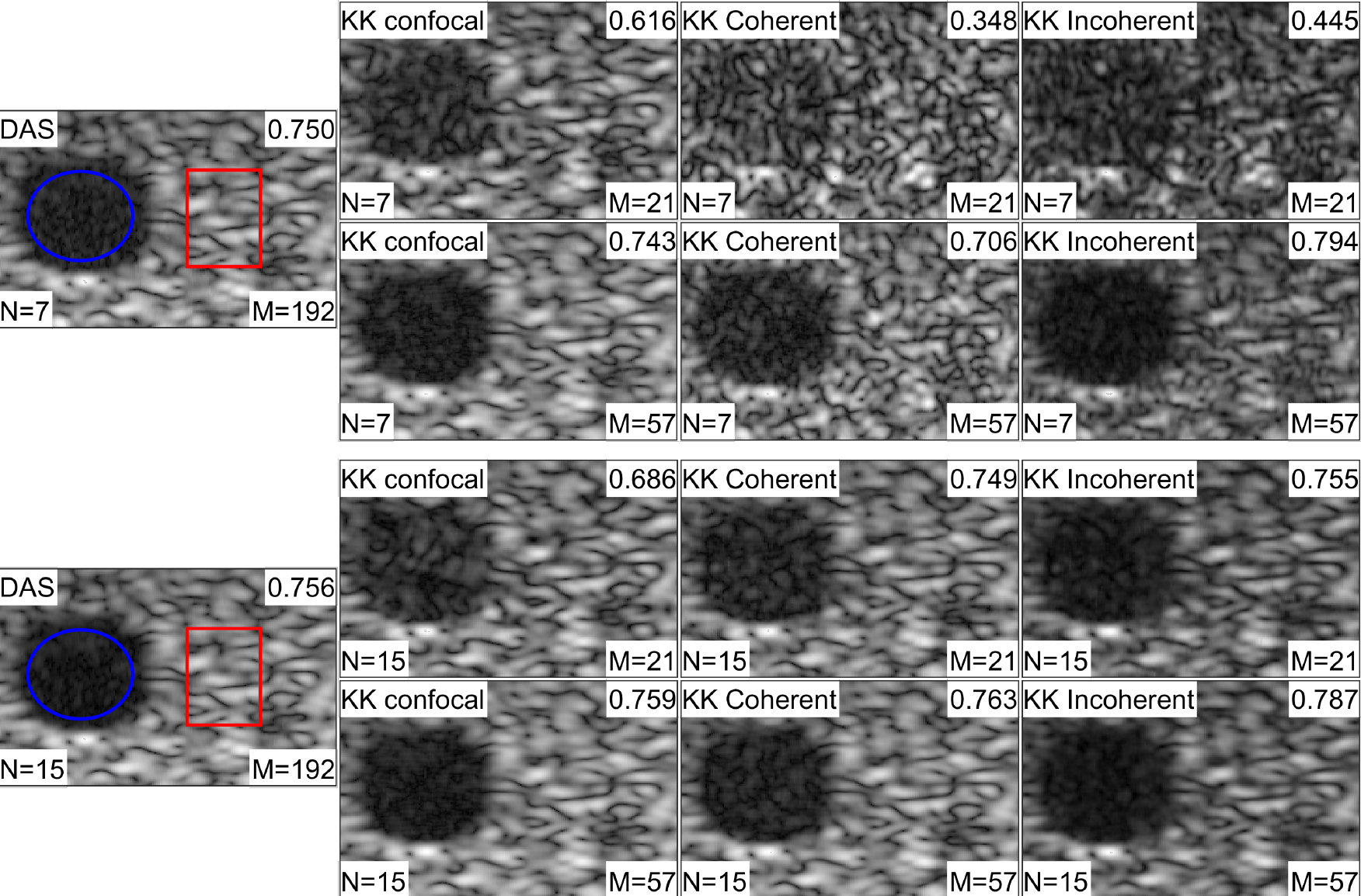}
\caption{Impact of reducing the number of transmit angles on image quality. Top half: $N=7$ transmit angles. Bottom half: $N=15$ transmit angles. In each half, the first row shows results obtained with more receive angles ($M=57$) and the second row with fewer angles ($M=21$). The DAS reference ($M=192$) is shown in left panels. The resultant gCNR values indicate that while reducing the number of transmit angles degrades contrast at high compression, increasing the number of receive angles can partially compensate for this loss, indicating a tradeoff between acquisition speed (dependent on $N$) and processing cost (dependent on $M$).}
\label{figAcq}
\end{figure*}

In Fig. \ref{figAcq} we evaluate the impact of reducing the total number of physical transmit angles $N$ when the number of synthesized receive angles $M$ is held fixed. Data was captured twice, with $N = 15$ and $N = 7$, while the probe was held at the same position. The overall range of transmit angles remained the same, to facilitate comparison. We observe that as the number of transmit angles decreases, so too does the resultant image quality, as expected and as quantified by the decrease in gCNR. On the other hand, the decrease in image quality that comes from reducing the number of transmit angles can be largely compensated for by increasing the number of synthesized receive angles. This too is expected from the well-known reciprocity linking transmit and receive directions\cite{rayleigh_application_1876}, and suggests an additional tradeoff between acquisition speed and processing speed. In other words, for applications where high frame rates are required, the number of physical acquisitions can be reduced and compensated for post-acquisition by beamforming with more synthesized receive angles.

\begin{table*}
    \caption{Processing Times}
    \label{tableTimes}
    \setlength{\tabcolsep}{4pt}
    \hspace*{20pt}
    \begin{tabular}{p{60pt}p{60pt}p{60pt}p{60pt}p{60pt}p{60pt}p{60pt}}
        
        Method N/M & ReorgAndFFT & Hilbert/Compress & IFFT & Beamform & Total & Comp. Ratio \\ 
        \hline 
        KK 7/7 & 6.8 & 3.4 & 0.5 & 2.6 & 15.8 & 27.4 \\ 
        KK 7/19 & 6.8 & 4.3 & 0.7 & 4.1 & 18.9 & 10.1 \\ 
        KK 7/21 & 6.8 & 4.6 & 0.8 & 4.1 & 19.2 & 9.1 \\ 
        KK 7/57 & 6.8 & 8.6 & 1.5 & 11.3 & 31.5 & 3.4 \\ 
        DAS 7/192 & 4.8 & 3.4 & 5 & 35.4 & 53.9 & 1 \\ 
        VS 7/192 & -- & -- & -- & -- & 18.3 & -- \\
        KK 15/7 & 13.3 & 6.6 & 0.7 & 4.7 & 30.7 & 27.4 \\ 
        KK 15/19 & 13.2 & 9 & 1.2 & 8 & 37.2 & 10.1 \\ 
        KK 15/21 & 13.3 & 9.5 & 1.2 & 8.2 & 38 & 9.1 \\ 
        KK 15/57 & 13.2 & 16.6 & 2.6 & 23.2 & 62.2 & 3.4 \\ 
        DAS 15/192 & 9.2 & 7 & 9.9 & 72.1 & 109.5 & 1 \\ 
        VS 15/192 & -- & -- & -- & -- & 38.7 & -- \\ 
        \hline\hline
        \multicolumn{6}{p{400pt}}{All units are in milliseconds. }
    \end{tabular}
\end{table*}

Table \ref{tableTimes} displays the processing speeds of the various custom EigenMex functions we used to perform KK and DAS beamforming. There are three key takeaways. First, KK beamforming does process faster than DAS beamforming in all configurations we tested. Second, the gain in speed is proportional to (but not equal to or greater than) the compression ratio. In other words, regardless of input data size, for a given compression ratio the gain in speed remains roughly the same. Third, Verasonics' optimized code is faster than our custom DAS implementation, suggesting there is room for improvement on the baseline performance of our KK beamforming implementations. 

\begin{figure*}[!]
\centering
\includegraphics[width= \textwidth]{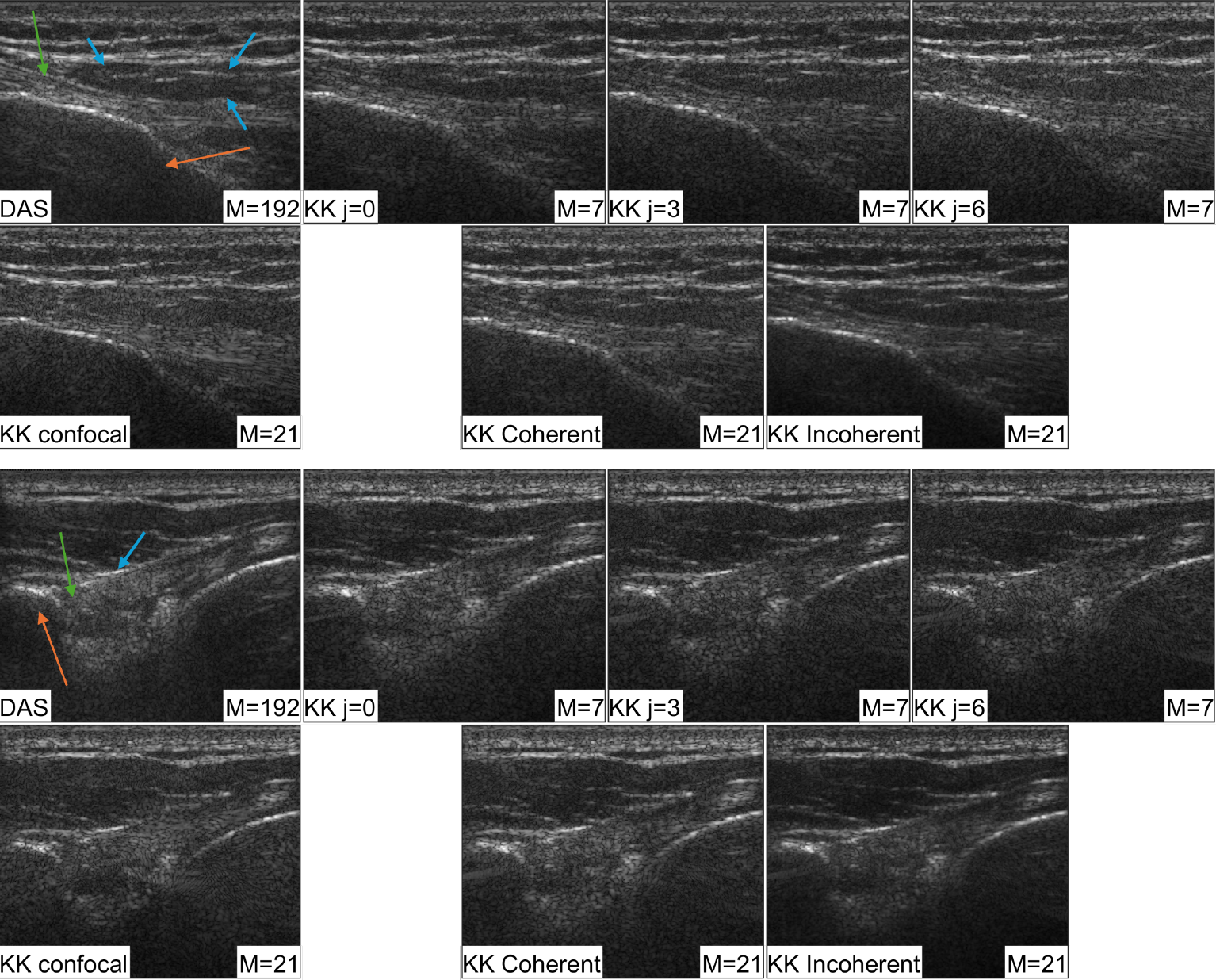}
\caption{KK beamforming of human tissue. Top two rows are images of hip bursa (green, orange and blue arrows indicate tendon, femur and fluid-filled bursa, respectively). Bottom two rows are images of acromioclavicular (AC) joint (green, orange and blue arrows indicate joint, acromion and clavicle, respectively). $M=7$ corresponds to $27 \times$ compression; $M=21$ corresponds to $9.1 \times$ compression. Coherent and incoherent panels are obtained by compounding $j=0,3,6$ panels directly above.}
\label{figBio}
\end{figure*}

Finally, we compare the performance of different KK beamforming implementations with conventional CPWC DAS when applied to human tissue imaging. Figure  \ref{figBio} shows annotated images of the hip bursa and acromioclavicular (AC) joint, where we note that such images are clinically relevant for the diagnosis of bursa inflammation (bursitis) or AC joint injury. As can be observed, KK beamforming provides only modest degradation in image quality even in the case of almost an order of magnitude reduction in RF data size. The hybrid implementation involving incoherent compounding, in particular, provides the best contrast of all, in agreement with results observed in Figs. \ref{figContrast} and \ref{figAcq}.

\section{DISCUSSION} 

KK beamforming is based on the decomposition of both the transmit and receive signals into plane waves. As shown above, this has many advantages. For one, it facilitates an interpretation of the beamforming process by transposing it fully into $k$-space. The roles of the transmit and receive signals become symmetric, and the interpretation of their respective contributions to image formation becomes straightforward. The spatial frequencies recovered from an object are given by simply $\mathbf{k}=\mathbf{k}_o -\mathbf{k}_i$, where KK beamforming offers direct control over both $\mathbf{k}_o$ and $\mathbf{k}_i$. Such direct control provides a large degree of flexibility in the manner in which the object spatial frequencies are sampled, opening the door to the optimization of sampling strategies for different purposes and under different constraints (computer processing power, memory limitations, speed requirements, etc.). 

As an example, we have shown that under the constraint of a restricted total number of samples, KK beamforming can distribute the samples more efficiently in $k$-space than conventional CPWC DAS, providing more uniform sampling with less sampling redundancy, thus leading to the possibility of data compression without significant loss of image quality. Importantly, KK beamforming provides a straightforward interpretation of the price paid for such data compression. In Eq. \ref{uniform} we provide a sampling strategy that offers direct control over the tradeoff between image resolution and contrast, where $j$ is the tuning parameter. For largest $j$ the $k$-space sampling becomes uniformly distributed over the largest range possible (broadest frequency support), providing a higher imaging spatial resolution than DAS even with fewer samples. However, this high resolution comes at a price. A broad frequency support spanned by a finite total number of samples unavoidably causes the sampling to become sparser, leading to reduced contrast resulting from an increased susceptibility to aliasing and clutter. By selecting smaller values of $j$, the frequency support becomes more constrained while continuing to maintain a mostly uniform sampling distribution. That is, the sampling becomes denser and contrast becomes increased, though at the cost of resolution. 

We note, however, that the sampling density provided by DAS is not uniform but rather it is denser at low spatial frequencies and sparser at high spatial frequencies. To mimic the performance of DAS, the sampling density provided by KK beamform can also be designed to follow this same general trend, which can be achieved in different ways. For example, it can be achieved directly by using a sampling distribution strategy prescribed by Eq. \ref{confocal}, referred to here as confocal (borrowing terminology from the optics community). As shown in Fig. \ref{fig3}c, the sampling density resulting from this is largely triangular in shape, centered about zero frequency, while avoiding sampling redundancies. Alternatively, one can continue to make use of the strategy prescribed by Eq. \ref{uniform}, though with multiple values of $j$ whose results are compounded. The compounding can be coherent (Eq. \ref{coherent}) or incoherent (Eq. \ref{incoherent}). In the examples provided in Figs. \ref{figRes}, \ref{figContrast}, \ref{figAcq} and \ref{figBio}, images were constructed from three values of $j$ ($j=0,3,6$). We found from these results that the hybrid approach involving the use of incoherent compounding is particularly effective at providing high image contrast, in some cases higher than DAS, even with fewer samples. Of course, different compounding strategies could be explored, for example making use of a larger number of $j$ values each with smaller values of $M$, or vice versa, which we will relegate to future work.

We have focused here on illustrating the benefits of KK beamforming for data compression. As shown in Table~\ref{tableTimes}, the processing time required for KK beamforming is consistently shorter than for our custom DAS implementation for all tested configurations. However, we do note that the measured gain in speed of approximately $1.7\times$ for the largest receive aperture ($M = 57$) falls somewhat short of the gain that might be expected from the  $3.4\times$ compression factor resulting from more cache-friendly memory management. This gap in speed performance is largely attributable to the prototype nature of our implementation: certain operations (notably Fourier transforms) were performed in MATLAB, with intermediate datasets copied between MATLAB and C++ at each stage. A fully integrated implementation with optimized memory management would likely close this gap. Nonetheless, our results demonstrate meaningful increases in speed even with a prototype software pipeline, with further gains to be expected from dedicated optimization or hardware acceleration. Our demonstration KK beamforming code is made freely available on our Github site (https://github.com/biomicroscopy/KKBeamforming). 

Additional future work can be envisaged. In our KK beamforming implementation, the compression step (Eq.~\ref{compression})  was performed in software post-acquisition. In principle, this step could be performed directly in hardware, for example making use of an ASIC \cite{chen_column-row-parallel_2016}, thus reducing both the computational cost and data bandwidth between probe and host computer. Such a hardware capability would be of particular interest for volumetric imaging with 2D arrays where actuator counts can reach into the thousands. 

In addition, the flexibility gained from direct control over $k$-space sampling can be exploited for the improvement of image quality by post-processing. For example, a variety of adaptive beamforming approaches have been developed to compensate for sample-induced aberrations or reduce image clutter/noise \cite{khetan_plane_2025,lambert_ultrasound_2022,kang_high-resolution_2017,agarwal_improving_2019,kou_high-resolution_2024,long_spatial_2022}. These approaches could be implemented fully in $k$-space and readily streamlined with the use of KK compression and beamforming. By exploiting the spatially-invariant nature of KK beamforming, the implementation of aberration correction by direct image deconvolution could also become facilitated. Again, we relegate the exploration of these possibilities to future work.

\section{Conclusion}

We have presented a symmetrized version of CPWC, called KK beamforming, wherein both transmit and receive signals are comprised of plane waves. KK beamforming can be implemented without any changes whatsoever to conventional CPWC RF acquisition protocols, and provides direct and flexible control over the spatial-frequency sampling distribution associated with image reconstruction. Our focus here was to demonstrate the effectiveness of KK beamforming for RF data compression. To this end, we presented different vernier-type sampling strategies to navigate the tradeoff between image resolution and contrast, and additionally provided strategies to approximate the frequency transfer function of conventional DAS beamforming while reducing sampling redundancy. Experimental results with calibration phantoms and human tissue demonstrate that KK beamforming preserves image quality comparable to DAS while enabling RF data compression by up to an order of magnitude, with corresponding reductions in processing times. KK beamforming is particularly attractive for high speed ultrasound imaging applications. 

\section*{Acknowledgements}

This work was partially supported by the National Science Foundation (EEC-2215990) and the National Institutes of Health (R21GM134216). The authors thank Alexandra Jureller from the Boston VA Orthopedic Surgery unit for help and guidance with human imaging examples.

\bibliographystyle{IEEEtran}
\bibliography{refZotero.bib, EigenRef.bib}
\end{document}